\documentclass[11pt]{article}
\usepackage[margin=1in]{geometry}
\usepackage{setspace}
\makeatletter
\def\endthebibliography{%
	\def\@noitemerr{\@latex@warning{Empty `thebibliography' environment}}%
	\endlist
}
\makeatother

\usepackage{hyperref}

\usepackage{amsfonts}
\usepackage{amsmath}
\usepackage{amssymb}
\usepackage{amsthm}
\usepackage{mathtools}
\usepackage{cases}

\usepackage{paralist}

\allowdisplaybreaks

\usepackage{verbatim}
\usepackage{graphicx}
\usepackage{subcaption}

\usepackage{xcolor}

\newtheorem{theorem}{Theorem}
\newtheorem{lemma}{Lemma}

\theoremstyle{definition}
\newtheorem{definition}{Definition}

\DeclareMathOperator*{\argmax}{arg\,max}

\newcommand{\xsol}{x^\mathrm{sol}}
\newcommand{\xopt}{x^\mathrm{opt}}

\title{Distributed Submodular Maximization with Parallel Execution}
\author{Haoyuan Sun, David Grimsman, and Jason R. Marden %
\thanks{H. Sun ({\tt hsun2@caltech.edu}) is with the California Institute of Technology (Caltech), Pasadena, CA. He was supported by Caltech's SURF program as the {\O}istein and Rita A. Skjellum SURF Fellow. He was also co-mentored by A. Wierman of Caltech.}
\thanks{D. Grimsman (\texttt{davidgrimsman@ucsb.edu}), J. R. Marden (\texttt{jmarden@ece.ucsb.edu}) are with the Department of Electrical and Computer Engineering, University of California, Santa Barbara, CA}
\thanks{This research was supported by NSF Grant \#ECCS-1638214 and U.S. Office of Naval Research (ONR) grant \#N00014-17-1-2060.}
\thanks{© 2020. This work has been accepted to American Control Conference for publication under a Creative Commons Licence CC-BY-NC-ND}}
\date{}

\begin{document}
	
\maketitle

\begin{abstract}
    The submodular maximization problem is widely applicable in many engineering problems where objectives exhibit diminishing returns. While this problem is known to be NP-hard for certain subclasses of objective functions, there is a greedy algorithm which guarantees approximation at least 1/2 of the optimal solution. This greedy algorithm can be implemented with a set of agents, each making a decision sequentially based on the choices of all prior agents. In this paper, we consider a generalization of the greedy algorithm in which agents can make decisions in parallel, rather than strictly in sequence. In particular, we are interested in partitioning the agents, where a set of agents in the partition all make a decision simultaneously based on the choices of prior agents, so that the algorithm terminates in limited iterations. We provide bounds on the performance of this parallelized version of the greedy algorithm and show that dividing the agents evenly among the sets in the partition yields an optimal structure. 
    It is shown that such optimal structures holds even under very relaxed information constraints. 
    We additionally show that this optimal structure is still near-optimal, even when additional information (i.e., total curvature) is known about the objective function. 
\end{abstract}

\clearpage

\section{Introduction}

Submodular maximization is an important topic with relevance to many fields and applications, including sensor placement \cite{krause2006near}, outbreak detection in networks \cite{leskovec2007cost}, maximizing and inferring influence in a social network \cite{kempe2003maximizing,gomez2012inferring}, document summarization \cite{lin2011class}, clustering \cite{mirzasoleiman2013distributed}, assigning satellites to targets \cite{qu2019distributed}, path planning for multiple robots \cite{singh2007efficient}, and leader selection and resource allocation in multiagent systems \cite{clark2011submodular,marden2016}.
An important similarity among these applications is the presence of an objective function which exhibits a ``diminishing returns" property. For instance, a group of leaders can impose some influence on a social network, but the marginal gain in influence achieved by adding a new leader to the group decreases as the size of the group increases. Objective functions (such as influence) satisfying this property are \emph{submodular}.

The submodular maximization problem is to choose a set of elements (such as social leaders) which maximize the submodular objective function, according to some constraints. This problem is known to be NP-hard for certain subclasses of submodular functions \cite{lovasz1983submodular}. Therefore, much research has focused on how to approximate the optimal solution \cite{calinescu2007maximizing,minoux1978accelerated,buchbinder2015tight,vondrak2008optimal}. The overall message of this research is that simple algorithms can perform well by providing solutions which are guaranteed to be within some factor of optimal.

One such algorithm is the greedy algorithm, first proposed in \cite{nemhauser1978analysis}. It was shown in this seminal work that for certain classes of constraints the solution provided by the greedy algorithm is guaranteed to be within $1-1/e$ of the optimal, and within $1/2$ of the optimal for the more general cases \cite{fisher1978analysis}. Since then, more sophisticated algorithms have been developed to show that there are many instances of the submodular maximization problem which can be solved efficiently within the $1-1/e$ guarantee \cite{calinescu2007maximizing,gairing2009covering}. It has also been shown that progress beyond this level of optimality is not possible using a polynomial-time algorithm \cite{feige1998threshold}.

More recent research has focused on distributed algorithms, since in many cases having a centralized agent with access to all the relevant data is untenable \cite{clark2016submodularity, mirzasoleiman2013distributed,corah2018distributed}. In this case, the greedy algorithm can be generalized using a set of $n$ agents, each with its own decision set. The combined set of decisions by the agents is evaluated by the submodular objective function, which they seek to maximize. Each agent chooses sequentially, maximizing its marginal contribution with respect to the decisions of the prior agents. In this setting, the greedy algorithm has been shown to provide a solution within $1/2$ of the optimal.
While simplistic from an algorithmic perspective, the informational requirement can be quite demanding as agents are required to observe all previous choices.
Accordingly, researchers have sought to characterize the impact of reducing this informational dependencies on the resulting performance guarantees~\cite{lin2011class,pan2014parallel,ene2018parallel}.

In this paper, we consider the case where no such centralized authority is present, i.e., the agents' decision sets are determined a priori and cannot be modified. Terminating the greedy algorithm in $q < n$ iterations thus requires a partitioning of the agents into sets $1, \dots, q$, where now each agent in set $j$ simultaneously chooses an action which maximizes its marginal contribution relative to the actions chosen by all the agents in sets $1, \dots, j-1$. In this setting, one can ask the following questions:
\begin{itemize}
    \item What is the best way to partition the agents? Should the agents be spread evenly across the sets in the partition, or should set 1 or set $q$ be larger than the others?
    \item Can we further reduce the amount of informational dependencies without sacrificing any performance guarantees?
    \item If some additional structure is known about the submodular objective function, how does that affect the partitioning strategy? 
\end{itemize}

In response to the questions listed above, the contributions of this paper are the following:
\begin{enumerate}
    \item Theorem \ref{thm:best-parallel} shows that given time constraint $q$, partitioning the agents into equally-sized sets is the best strategy to parallelize the greedy algorithm.
    \item Theorem \ref{thm:best-parallel-graph} proves that we can relax the information sharing constraints under a graph-theoretic interpretation and still achieve the same performance guarantees.
    \item Theorem \ref{thm:best-parallel-strict-monotone-graph} shows that if we know some additional properties of the submodular function, i.e., their total curvature, then we have improved performance guarantees and the strategy in Theorem \ref{thm:best-parallel-graph} is nearly optimal.
\end{enumerate}

This paper is organized as follows: 
\begin{enumerate}
    \item Section \ref{sec:model} introduces general definitions and the greedy algorithm;
    \item Section \ref{sec:parallel-computation} introduces Theorem \ref{thm:best-parallel};
    \item Section \ref{sec:parallel-communication} gives a more general graph-theoretic interpretation of the model and introduces Theorem \ref{thm:best-parallel-graph}.
    \item Section \ref{sec:beta-strict-monotone} defines the total curvature property and introduces Theorem \ref{thm:best-parallel-strict-monotone-graph}.
\end{enumerate}

\section{Model}
\label{sec:model}

\subsection{Distributed Submodular Optimization}
Consider a base set $S$ and a set function $f : 2^S \to \mathbb{R}_{\ge 0}$.
Given $A, B \subseteq S$, we define the \textit{marginal contribution} of $A$ with respect to $B$ as 
\begin{equation}
f(A \mid B) = f(A \cup B) - f(B) 
\end{equation}

In this paper, we are interested in distributed algorithms for maximization of submodular functions, where there is a collection of $n$ decision-making agents, named as $N = \{1, \dots, n\}$, and a set of decisions $S$.
Each decision-making agent is associated with a set of permissible decisions $X_i \subseteq S$ so that $X_1, \dots, X_n$ form a partition of $S$.
For rest of the paper, we refer to decision-making agents as simply agents.
Then the collection of decisions by all agents $x = \{x_1, \dots, x_n\}$ is called an \textit{action profile}, and we denote the set of all action profiles as $X = X_1 \times \cdots \times X_n$.
An action profile $x \in X$ is evaluated by an \textit{objective function} $f : 2^S \to \mathbb{R}_{\ge 0}$ as $f(x) = f(\cup_i \{x_i\})$.
Furthermore, we restrict our attention to $f$'s which satisfy the following properties:
\begin{itemize}
    \item \textbf{Normalized} $f(\emptyset) = 0$.
    \item \textbf{Monotonic} $f(\{e\} \mid A) \ge 0$ for all $e \in S$ and $A \subseteq S$.
    \item \textbf{Submodular} $f(\{e\} \mid A) \ge f(\{e\} \mid B)$ for all $A \subseteq B \subseteq S$ and $e \in S \setminus B$.
\end{itemize}
For simplicity, define $\mathcal{F}$ as the set of functions satisfying these three properties.
For any $f \in \mathcal{F}$, the goal of the submodular maximization problem is to find:
\begin{equation}
\label{def:submodular-maximization}
    \xopt \in \argmax_{x \in X} f(x)
\end{equation}

For convenience, we overload $f$ with the understanding that $f(e) = f(\{e\})$ for $e \in S$, and $f(A, B) = f(A \cup B)$ for $A, B \subseteq S$.
We also introduction the follow notation.
For any $j, k \in \{1, \dots, n\}, j < k$, let $[k] = \{1, \dots, k\}$ and $j : k = \{j, i+1, \dots, k\}$.
Furthermore, for any action profile $x \in X$ and set $M \subseteq N$, let $f(x_M) = f(\cup_{i \in M} \{x_i\})$.
For instance, $f(x_{1:10}) = f(\{x_1, \dots, x_{10}\})$.

\begin{figure}
\centering
\begin{subfigure}{0.8\textwidth}
    \centering
    \includegraphics[width=0.8\textwidth]{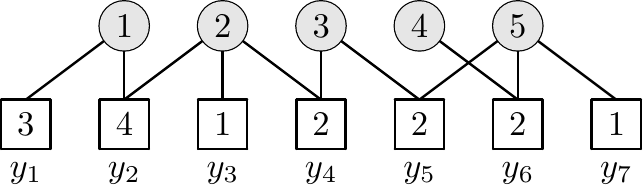}
    \caption{Illustration of weighted subset cover problem. The 5 agents are represented by circles. The target set $Y = \{y_1, \dots y_7\}$ is represented by boxes.
    For each box, its label indicates its weight, and the black lines indicate the permissible decisions, for example, $T(X_1) = \{\{y_1\}, \{y_2\}\}$.}
\end{subfigure}

\begin{subfigure}{0.9\textwidth}
    \centering
    \begin{tabular}{l|c|c|c|c|c|c}
             Algorithm &  $x_1$ & $x_2$ & $x_3$ & $x_4$ & $x_5$ & $f(x_{1:5})$ \\
             \hline
             \hline
             Optimal & $y_1$ & $y_2$ & $y_4$ & $y_6$ & $y_5$ & 13 \\
             \hline
             Greedy & $y_2$ & $y_4$ & $y_5$ & $y_6$ & $y_7$ & 11 \\
             \hline
             Parallel ($P_1$) & $y_2$ & $y_2$ & $y_5$ & $y_6$ & $y_6$ & 8 \\
             \hline
             Parallel ($P_2$) & $y_2$ & $y_2$ & $y_5$ & $y_6$ & $y_7$ & 9 \\
    \end{tabular}

    \caption{For the above weighted set cover problem, this table shows the decisions $x_1, \dots, x_5$ by several algorithms described in this paper.
    $P_1$ and $P_2$ are the optimal iteration assignments as shown in Fig. \ref{fig:weird-parallel} and \ref{fig:standard-parallel}, respectively.
    For simplicity, we used element instead of set to denote each decision, e.g. we wrote $y_1$ but we mean $\{y_1\}$.
    To illustrate how the greedy algorithm works, let's consider $P_2$.
    We have $x_1 = x_2 = \{y_2\}$ because agents 1 and 2 are deciding at the same time.
    Then agents 3 and 4 choose at the same time, and one possible result is $x_3 = \{y_5\}, x_4 = \{y_6\}$.
    Lastly, agent 5 makes a decision, it sees that agents 3 and 4 already took $y_5$ and $y_6$, so it chooses $\{y_7\}$ as the decision maximizing its marginal contribution.}
    \label{table:run}
\end{subfigure}

\caption{An instance of weighted subset cover problem and the behavior of various algorithms for this problem.}
\label{fig:weighted-subset-cover}
\end{figure}

An example of the submodular maximization problem is the weighted set cover problem~\cite{gairing2009covering}.
Given a target set $Y$ and a mapping $T : S \to 2^Y$ from the decisions to subsets of $Y$,\footnote{
Note that the decisions are related to assignments to the subsets of $Y$ in the sense that each decision is an ordered pair (agent ID, subset of $Y$).
This way we allow agents to choose the same element from $Y$ while maintaining that the decisions are distinct.
} 
the value of an action profile $x \in X$ is determined by a weight function $w : Y \to \mathbb{R}_{\ge 0}$:
\begin{equation}
\label{def:weighted-subset}
    f(x) = \sum_{y \in \cup_i T(x_i)} w(y)
\end{equation}

Intuitively, this problem aims to ``cover'' as much of the target set as possible.
An instance of this problem is illustrated by Figure \ref{fig:weighted-subset-cover}.


\subsection{Greedy Algorithm and Parallelization}

A distributed greedy algorithm can be used to approximate the problem as stated in \eqref{def:submodular-maximization}.
In the greedy algorithm, the agents make decision in sequential order according to their names $i \in N$ and each agent attempts to maximize its marginal contribution with respect to the choices of prior agents:
\begin{equation}
\label{def:greedy-algorithm}
    \xsol_i \in \argmax_{x_i \in X_i} f(x_i \mid x_{1:i-1})
\end{equation}
Note that the greedy solution $\xsol$ may not be unique.
In the context of this paper, we assume that $\xsol$ is the worst solution among all possible greedy solutions.
Define $\gamma(f, X)$ to be $f(\xsol) / f(\xopt)$ from the greedy algorithm subject to objective function $f$ and set of action profiles $X$.
Then, to measure the quality of the greedy algorithm, we consider its \textit{competitive ratio}:
\begin{equation}
    \gamma = \inf_{f \in \mathcal{F}, X} \gamma(f, X)
\end{equation}

The well known result from \cite{fisher1978analysis} states that $\gamma = 1/2$, which means that \eqref{def:greedy-algorithm} guarantees that the performance of the greedy solution is always within 1/2 of optimal.  
Furthermore, only one agent can make a decision at any given time and thus a solution will be found in precisely $n$ iterations. 

The focus of this work is on characterizing the achievable competitive ratio for situations where a system-design does not have the luxury of having $n$ iterations to construct a decision.
To this end, we introduce the parallelized greedy algorithm to allow multiple agents to make decisions at the same time.
More formally, we consider a situation where the system is given limited number of iterations $q \leq n$ and needs to come up with an iteration assignment $P : [n] \to [q]$ so that each agent makes decision only based off the choices made by agents from earlier iterations:
\begin{equation}
\label{def:parallel-greedy}
    \xsol_i \in \argmax_{x_i \in X_i} f(x_i \mid x_{\mathcal{N}_i})
\end{equation}
where $\mathcal{N}_i = \{j \mid P(j) < P(i)\} \subseteq \{1, \dots, i-1\}$.
Note, that to maintain consistency with \eqref{def:greedy-algorithm}, we require that $P$ preserves the implicit ordering of the greedy algorithm: $P(i) \le P(j)$ whenever $i < j$.

We are interested in the performance of \eqref{def:parallel-greedy} given a fixed number of agents and number of iterations.
We define the competitive ratio of a time step assignment $P$ as:
\begin{equation}
\label{def:comp-ratio-parallel}
    \gamma(P) = \inf_{f \in \mathcal{F}, X} \gamma(f, X, P)
\end{equation}
where $\gamma(f, X, P)$ is $f(\xsol) / f(\xopt)$ from the parallelized greedy algorithm subject to objective function $f$, set of action profiles $X$ and iteration assignment $P$.
Denote the set of all possible iteration assignments with $n$ agents and at most $q$ iterations as: \begin{equation}
\mathcal{P}_{n, q} = \{P : [n] \to [q] \}
\end{equation}
We seek to find the best possible competitive ratio, and the iteration assignment which achieves the optimal competitive ratio (if such an assignment exists):
\begin{equation}
\label{def:comp-ratio-over-all-parallel}
    \rho(n, q) = \sup_{P \in \mathcal{P}_{n, q}} \gamma(P)
\end{equation}

\begin{equation}
\label{def:optimal-parallel}
    P^*_{n, q} \in \argmax_{P \in \mathcal{P}_{n, q}} \gamma(P)
\end{equation}


\section{Optimal Parallel Structures}
\label{sec:parallel-computation}

In this section, we present the best possible competitive ratio for the parallelized greedy algorithm and optimal iteration assignment which achieves such a bound.

Note that according to \eqref{def:parallel-greedy}, a pair of agents do not utilize each other's decisions (in either direction) when they are in the same iteration.
This intuitively represent some ``blindspots'' in the parallelized greedy algorithm compared to the original greedy algorithm.
One way to reduce those ``blindspots'' is to divide the agents evenly among the available iterations, in other words, we want the number of agents deciding in parallel to be as close to the average number as possible.
Theorem \ref{thm:best-parallel} illustrates that this simple idea is nearly sufficient to yield the best possible iteration assignment.

\begin{theorem}
\label{thm:best-parallel}
    Given a parallelized submodular maximization problem with $n$ agents and $q$ iterations, the competitive ratio is:
    \begin{equation}
    \label{equ:parallel-exact}
        \rho(n, q) =
        \begin{cases}
            \frac{1}{r} & \text{if } n \equiv 1 \pmod{q} \\
            \frac{1}{r+1} & \text{otherwise} \\
        \end{cases}
    \end{equation}
    where $r = \lceil n / q \rceil$. In particular, when $n \equiv 1 \pmod{q}$,
    \begin{equation}
    \label{equ:weird-parallel}
        P^*_{n, q} =
        \begin{cases}
            \left\lceil \frac{i}{r-1} \right\rceil & \text{if } i < n \\
            q & \text{if } i = n
        \end{cases}
    \end{equation}
    Otherwise,
    \begin{equation}
    \label{equ:standard-parallel}
        P^*_{n, q} = \left\lceil \frac{i}{r} \right\rceil
    \end{equation}
\end{theorem}

\begin{figure*}
\begin{subfigure}{0.20\textwidth}
    \centering
    \includegraphics[scale=1.2]{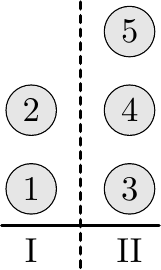}
    \caption{The optimal assignment for $n = 5$, $q  = 2$ as described in \eqref{equ:weird-parallel}.}
    \label{fig:weird-parallel}
\end{subfigure}
\hfill
\begin{subfigure}{0.20\textwidth}
    \centering
    \includegraphics[scale=1.2]{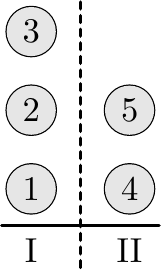}
    \caption{A non-optimal assignment for $n=5, q = 2$.}
    \label{fig:bad-parallel-weird}
\end{subfigure}
\hfill
\begin{subfigure}{0.25\textwidth}
    \centering
    \includegraphics[scale=1.2]{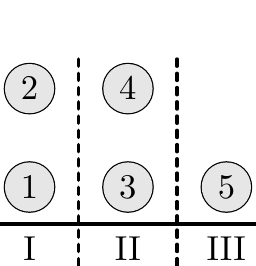}
    \caption{The optimal assignment for $n = 5$, $q  = 3$ as described in \eqref{equ:standard-parallel}.}
    \label{fig:standard-parallel}
\end{subfigure}
\hfill
\begin{subfigure}{0.27\textwidth}
    \centering
    \includegraphics[scale=1.2]{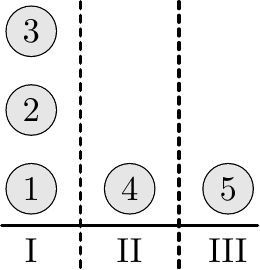}
    \caption{A non-optimal assignment for $n=5, q = 3$.}
    \label{fig:bad-parallel}
\end{subfigure}
\caption{Examples of optimal iteration assignments as given by Theorem \ref{thm:best-parallel}. 
Each agent, represented by a node, is named as $1, \dots, n$. Each column, labeled by a Roman numeral, contains all agents executed at a given iteration.
Also shown are some examples of iteration assignments that are not optimal.
According to Theorem \ref{thm:best-parallel}, the competitive ratio in Fig. \ref{fig:weird-parallel} is $1/3$ and the competitive ratio in Fig. \ref{fig:standard-parallel} is also $1/3$.
On the other hand, in Section \ref{sec:proof-parallel}, Lemma \ref{thm:david} shows that the competitive ratio in Fig. \ref{fig:bad-parallel-weird} and Fig. \ref{fig:bad-parallel} are both $1/4$.
}
\label{fig:parallel}

\end{figure*}

Figure \ref{fig:parallel} illustrates some example of optimal and non-optimal assignments.
In particular, Figure \ref{fig:bad-parallel-weird} gives an edge case where our simple intuition is not enough to make the optimal assignment.
And Figure \ref{fig:bad-parallel} shows that other approaches, such as assigning more agents to earlier iterations to reduce the ``blindness" in later iterations, are not optimal.

We observe that the competitive ratio is roughly inversely proportional to the value $r$.
Theorem \ref{thm:best-parallel} as stated does not offer a good explanation for why this relation holds.
To better understand this, we will generalize our model in Section \ref{sec:parallel-communication} and present a strengthening of Theorem \ref{thm:best-parallel} in Section \ref{sec:proof-parallel}.

\section{Parallelization as Information Exchange}
\label{sec:communication-graph}

We observe that our parallelization approach implicitly defines a certain type of informational dependencies.
In both \eqref{def:greedy-algorithm} and \eqref{def:parallel-greedy}, each agent uses the choices of prior agents to make its own decision.
Alternatively, we can view this process as an agent communicating its decision to other agents who depend on this piece of information.
In our parallelization model, each agent is required to broadcast its decision to all of the later agents.
This requirements seems rather restrictive and thus in this section, we will reduce the amount of informational dependencies by relaxing the constraints.
Specifically, we will treat the communication between one pair of agents as an individual entity so that an agent only has to send its choice to selected later agents and receive choices from selected previous agents.

\subsection{Parallelization and Information}
\label{sec:parallel-communication}

\begin{figure*}
\begin{subfigure}{0.50\textwidth}
    \centering
    \includegraphics[scale=1.2]{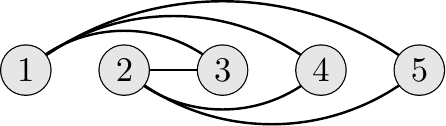}
    \caption{The optimal assignment for $n = 5$, $q  = 2$ as described in \eqref{equ:weird-parallel}.}
    \label{fig:weird-parallel-graph}
\end{subfigure}
\hfill
\begin{subfigure}{0.45\textwidth}
    \centering
    \includegraphics[scale=1.2]{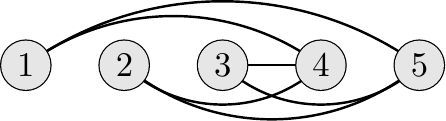}
    \caption{A non-optimal assignment for $n=5, q = 2$.}
    \label{fig:bad-parallel-weird-graph}
\end{subfigure}
\hfill
\begin{subfigure}{0.50\textwidth}
    \centering
    \includegraphics[scale=1.2]{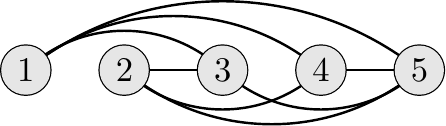}
    \caption{The optimal assignment for $n = 5$, $q  = 3$ as described in \eqref{equ:standard-parallel}.}
    \label{fig:standard-parallel-graph}
\end{subfigure}
\hfill
\begin{subfigure}{0.45\textwidth}
    \centering
    \includegraphics[scale=1.2]{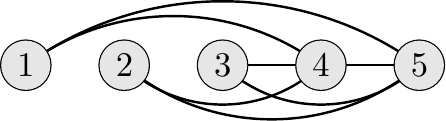}
    \caption{A non-optimal assignment for $n=5, q = 3$.}
    \label{fig:bad-parallel-graph}
\end{subfigure}

\caption{
The induced information graph of iteration assignments from Figure \ref{fig:parallel} as discussed in Section \ref{sec:communication-graph}.
In the information graphs, an edge $(i, j)$ where $i < j$ represents that the iteration of agent $i$ is before the iteration of agent $j$ and hence agent $j$ requires knowing the choice made by agent $j$ before making its own decision.
Each edge can be thought of as the flow of information exchanges between the agents in which a prior agent share its choice so later agents can decide.
}
\label{fig:parallel-graph}

\end{figure*}

We can model the information exchange as an undirected graph $G = (V, E)$, where $V = N$ represents the agents and each edge $(i, j) \in E$ represents information being exchanged between two agents.
Since the greedy algorithm has an implicit ordering induced by the agents' names, for every pair $i < j$ so that $(i, j) \in E$, agent $j$ requires knowing the choice of agent $i$, but agents $i$ does not access agent $j$'s decision.
Hence, we can use $G$, which we will call the \textit{information graph}, to determine the set of prior agents which agent $i \in N$ accesses as $\mathcal{N}_i = \{j < i \mid (j, i) \in E\} \subseteq [i-1]$.
From here, it is natural to consider the generalized greedy algorithm proposed in \cite{gharesifard2017distributed}:
\begin{equation}
\label{def:general-greedy}
    \xsol_i \in \argmax_{x_i \in X_i} f(x_i \mid x_{\mathcal{N}_i})
\end{equation}

First we noticed that the parallelized greedy algorithm as defined in \eqref{def:parallel-greedy} is a special case of \eqref{def:general-greedy}.
In the parallelized greedy algorithm, each agent requires information from all agents executed in prior iterations.
Therefore, an iteration assignment $P \in \mathcal{P}_{n, q}$ induces a corresponding information graph $G = (V, E)$ so that $V = N$, $E = \{(i, j) \mid P(i) < P(j)\}$.
Figure \ref{fig:parallel-graph} illustrates how the iteration assignments from Figure \ref{fig:parallel} induce corresponding information graphs.
For rest of this section, we mention an iteration assignment with the understanding that we are interested in its induced information graph.

Conversely, an information graph induces an ordering to parallelize the generalized greedy algorithm.
If all agents in $\mathcal{N}_i$ have made their decisions but agent $j < i$ is still undecided, then we can let agents $i$ and $j$ to make their decisions in the same iteration without affecting the algorithm.
Intuitively, we can think of information being relayed through some paths in $G$ and the earliest iteration for which agent $i$ can decide depends on the longest path in $G$ leading to node $i$.
Formally, a simple induction argument shows that the function $\overline{P} : G \times [n] \to [n]$ determines the earliest iteration in which an agent could decide:
\begin{equation}
\overline{P}(G; i) = 
\begin{cases}
1 & \text{if } \mathcal{N}_i = \emptyset \\
1 + \max_{j \in \mathcal{N}_i} \overline{P}(G; j) & \text{otherwise}
\end{cases}
\end{equation}
This function $\overline{P}$ can be used to construct a parallelization of the greedy algorithm in which agent $i$ makes its decision in iteration $\overline{P}(G; i)$ and for every $j \in \mathcal{N}_i$, $\overline{P}(G; j) < \overline{P}(G; i)$.
From here we can define $\mathcal{G}_{n, q}$ as the set of graphs which induces a parallelization with at most $q$ iterations:
\begin{equation}
    \mathcal{G}_{n, q} = \{G = (V, E) : |V| = n, \max_i \overline{P}(G; i) \le q\}
\end{equation}
From the the definitions, it is clear that $\mathcal{P}_{n, q} \subseteq \mathcal{G}_{n, q}$. 
Hence we can adopt competitive ratios for information graphs so that \eqref{def:comp-ratio-parallel}, \eqref{def:comp-ratio-over-all-parallel} and \eqref{def:optimal-parallel} become:
\begin{equation}
\label{def:comp-ratio-graph}
        \gamma(G) = \inf_{f, X} \gamma(f, X, G)
\end{equation}
\begin{equation}
\label{def:comp-ratio-over-all-graph}
    \eta(n, q) = \sup_{G \in \mathcal{G}_{n, q}} \gamma(G)
\end{equation}
\begin{equation}
\label{def:optimal-graph}
    G^*_{n, q} \in \argmax_{G \in \mathcal{G}_{n, q}} \gamma(G)
\end{equation}

where $\gamma(f, X, G)$ is $f(\xsol) / f(\xopt)$ from the generalized greedy algorithm subject to objective function $f$, set of action profiles $X$ and information graph $G$.

It is natural to ask whether the above generalization of parallelized greedy algorithm yields higher competitive ratio under the same number of agents and iterations.
Also, in some applications, the information exchange may incur some costs, and therefore having a information graph with many edges is undesirable.
So we wish to answer a second question, whether we can achieve the same competitive ratio in \eqref{equ:parallel-exact} 
with information graphs that have fewer edges than that of \eqref{equ:weird-parallel} and \eqref{equ:standard-parallel}.
Theorems \ref{thm:best-parallel-graph},
as presented below, show that generalizing to information graphs does not yield higher competitive ratio but allows us to achieve the same optimal competitive ratio with fewer edges.

\subsection{Preliminary: Graph Theory}
The main result on information graphs leverages several concepts from graph theory.
We assume that we have an undirected graph $G = (V, E)$:
\begin{definition}
Nodes $K \subseteq V$ form a \textit{clique} if for all distinct $i, j \in K$, $(i, j) \in E$.
The \textit{clique number} $\omega(G)$ of a graph $G$ is the size of the maximum clique in $G$.
\end{definition}

\begin{definition}
A \textit{clique cover} of a graph $G$ is a partition of $V$ so that each set in the partition forms a clique.
And the \textit{clique cover number} $\theta(G)$ is the least number of cliques necessary to form a clique cover.
\end{definition}

\begin{definition}
Nodes $I \subseteq V$ form an \textit{independent set} if for all distinct $i, j \in I$, $(i, j) \not\in E$.
The \textit{maximum independent set} is an independent set with the largest possible number of nodes.
And the \textit{independence number} $\alpha(G)$ is the size of the maximum independent set.
\end{definition}

Lastly, a \textit{Tur\'{a}n graph} $T(n, r)$ of $n$ vertices and clique number $r$ is constructed as follows:
\begin{itemize}
    \item We partition the $n$ vertices into $r$ subsets as evenly as possible. More specifically, we have $(n \mod{r})$ subsets with $\lceil n/r \rceil$ vertices and $(r - n \mod{r})$ subsets with $\lfloor n/r \rfloor$ vertices.
    \item We draw an edge between two vertices if and only if they belong to different subsets.
\end{itemize}
And it has the following interesting property~\cite{turan1941external}:
\begin{lemma}[Tur\'{a}n]
    The Tur\'{a}n graph $T(n, r)$ is the $n$-vertex graph with the largest number of edges that does not contain an $(r+1)$-clique.
    In other words:
    \begin{equation}
        T(n, r) = \argmax_{G = (V, E): \, |V| = n, \, \omega(G) \le r} |E|
    \end{equation}
\end{lemma}


\subsection{Main Result on Information Graphs}
\label{sec:proof-parallel}

We will present a stronger version of Theorem \ref{thm:best-parallel} that generalizes to the setting of information graphs.

\begin{theorem}
\label{thm:best-parallel-graph}
    Given a parallelized submodular maximization problem with $n$ agents and $q$ iterations, the competitive ratio is (recall that $r = \lceil n / q \rceil$):
    \begin{equation}
    \label{equ:parallel-exact-graph}
        \eta(n, q) =
        \begin{cases}
            \frac{1}{r} & \text{if } n \equiv 1 \pmod{q} \\
            \frac{1}{r + 1} & \text{otherwise} \\
        \end{cases}
    \end{equation}
    In particular, when $n \equiv 1 \pmod{q}$, for the following edge set $E$, $G = (V, E)$ achieves the equality $\gamma(G) = 1/r$:
    \begin{equation}
    \label{equ:weird-graph}
        E = \left\{
        \begin{aligned}
        (i, j) & \text{ for all } i < j < n, i \equiv j \, (\mathrm{mod} \; r-1) \\
        (i, n) & \text{ for all } i\le (q-1)(r-1)
        \end{aligned}
        \right\}
    \end{equation}
    Otherwise, for the following edge set $E$, $G = (V, E)$ achieves the equality $\gamma(G) = 1/(r+1)$:
    \begin{equation}
    \label{equ:complement-turan}
        E = \{(i, j) \mid i \neq j, i \equiv j \, (\mathrm{mod} \; r) \}
    \end{equation}
\end{theorem}

\begin{figure}
\begin{subfigure}{0.48\textwidth}
    \centering
    \includegraphics[scale=1.2]{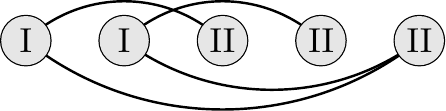}
    \caption{The optimal graph for $n = 5$, $q  = 2$ as described in \eqref{equ:weird-graph}.}
    \label{fig:weird-graph}
\end{subfigure}
~
\begin{subfigure}{0.48\textwidth}
    \centering
    \includegraphics[scale=1.2]{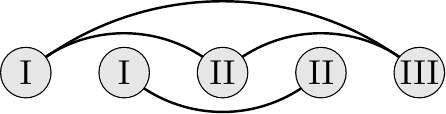}
    \caption{The optimal graph for $n = 5$, $q  = 3$ as described in \eqref{equ:complement-turan}.}
    \label{fig:complement-turan}
\end{subfigure}

\caption{Examples of optimal information graph as given by Theorem \ref{thm:best-parallel-graph}.
The agents, represented by nodes, are implicitly named as $1, \dots n$ from left to right and the Roman numeral labels indicate the iteration in which the agents will be executed. 
Note that Fig. \ref{fig:weird-graph} is a subgraph of Fig. \ref{fig:weird-parallel-graph} and Fig. \ref{fig:complement-turan} is a subgraph of Fig. \ref{fig:standard-parallel-graph}.
All four of these graphs achieve their respective optimal competitive ratio, which all happen to be $1/3$.
However, both Fig. \ref{fig:weird-graph} and Fig. \ref{fig:complement-turan} have 4 edges, whereas Fig. \ref{fig:weird-parallel-graph} has 6 edges and Fig. \ref{fig:standard-parallel-graph} has 8 edges.
Also, by Theorem \ref{thm:best-parallel-strict-monotone-graph},  when the objective function has total curvature $\lambda$, Fig. \ref{fig:weird-graph} and Fig. \ref{fig:complement-turan} are still nearly-optimal.
}
\label{fig:graph}

\end{figure}

Figure \ref{fig:graph} illustrates the aforementioned optimal information graphs.
Note that \eqref{equ:weird-graph} and \eqref{equ:complement-turan} are subgraphs of the information graph induced by \eqref{equ:weird-parallel} and \eqref{equ:standard-parallel}, respectively.
Furthermore, \eqref{equ:weird-graph} has approximately $1/(r-1)$ times as many edges as \eqref{equ:weird-parallel} and \eqref{equ:complement-turan} has approximately $1/r$ times as many edges as \eqref{equ:standard-parallel}.
This is true because every agent (except agent $n$ when $n \equiv 1 \pmod{q}$) requires information from just one agent from each prior time step.
Additonally, \eqref{equ:complement-turan} is ``delightfully parallel'' so that we can partition the agents into threads according to the clique in which they reside, and no information need to be exchanged among threads.

The following lemma shows that the value $r$ is related to the independence number of the information graph:
\begin{lemma}
\label{thm:parallel-indep-number}
For any $G \in \mathcal{G}_{n, q}$, $\alpha(G) \ge r = \lceil n / q \rceil$.
\end{lemma}

\begin{proof}
Let $D_k = \{i \mid \overline{P}(G; i) = k\}$ be the set of agents executed in the $k$th iteration.
From out construction, $\bigcup_{1 \le k \le q} D_k = [n]$.
By the pigeonhole principle, $\max_{1 \le k \le q} |D_k| \ge r$.
Since $D_k$ must be an independent set for any $1 \le k \le q$, we conclude that $\alpha(G) \ge r$.
\end{proof}

Since the agents in an independent set do not rely on each other to make a decision, they can be assigned to the same iteration.
Therefore $r$ is a lower bound on how many agents can decide in parallel.

The proof of Theorem \ref{thm:best-parallel-graph} relies on Theorem 1 and Corollary 1 from~\cite{grimsman2018impact} that relates the competitive ratio of an information graph to its independence number and clique cover number:
\begin{lemma}
\label{thm:david}
Given any information graph $G = (V, E)$,
\begin{equation}
    \frac{1}{\alpha(G)} \ge \gamma(G) \ge \frac{1}{\theta(G) + 1} 
\end{equation}
\end{lemma}

\begin{lemma}
\label{thm:david-sibling}
If there exists $w \in V$ and a maximum independent set $I$ s.t. $i \in \mathcal{N}_w$ for some $i \in I$, then
\begin{equation}
    \gamma(G) \le \frac{1}{\alpha(G) + 1}
\end{equation}
\end{lemma}

By combining Lemmas \ref{thm:parallel-indep-number} and \ref{thm:david}, we can see that, intuitively, the competitive ratio $\eta$ should be approximately $1/r$.
Lastly, we note that the graph described by \eqref{equ:complement-turan} is in fact a complement Tur\`{a}n graph.
This reinforces our earlier intuitions and relates to Lemma \ref{thm:david} that the optimal parallel structure arises from having as few agents decide in parallel as possible.
For the full proof of Theorem \ref{thm:best-parallel-graph}, please refer to Appendix \ref{sec:complete-proof-parallel}.




\section{Submodular Functions with Total Curvature}
\label{sec:beta-strict-monotone}

Recent research, such as \cite{corah2018distributed}, has been focused on whether imposing additional structure on the objective function and the set of action profiles would help the underlying greedy algorithm to yield better performing solutions.
One intuition is to consider the scenario when actions do not ``overlap'' in the sense that each decision is guaranteed to make some marginal contribution.
Here, we will analyze how the property introduced in \cite{conforti1984submodular} enables a parallelized greedy algorithm to yield higher competitive ratio.

\begin{figure}
    \centering
    \includegraphics[scale=0.5]{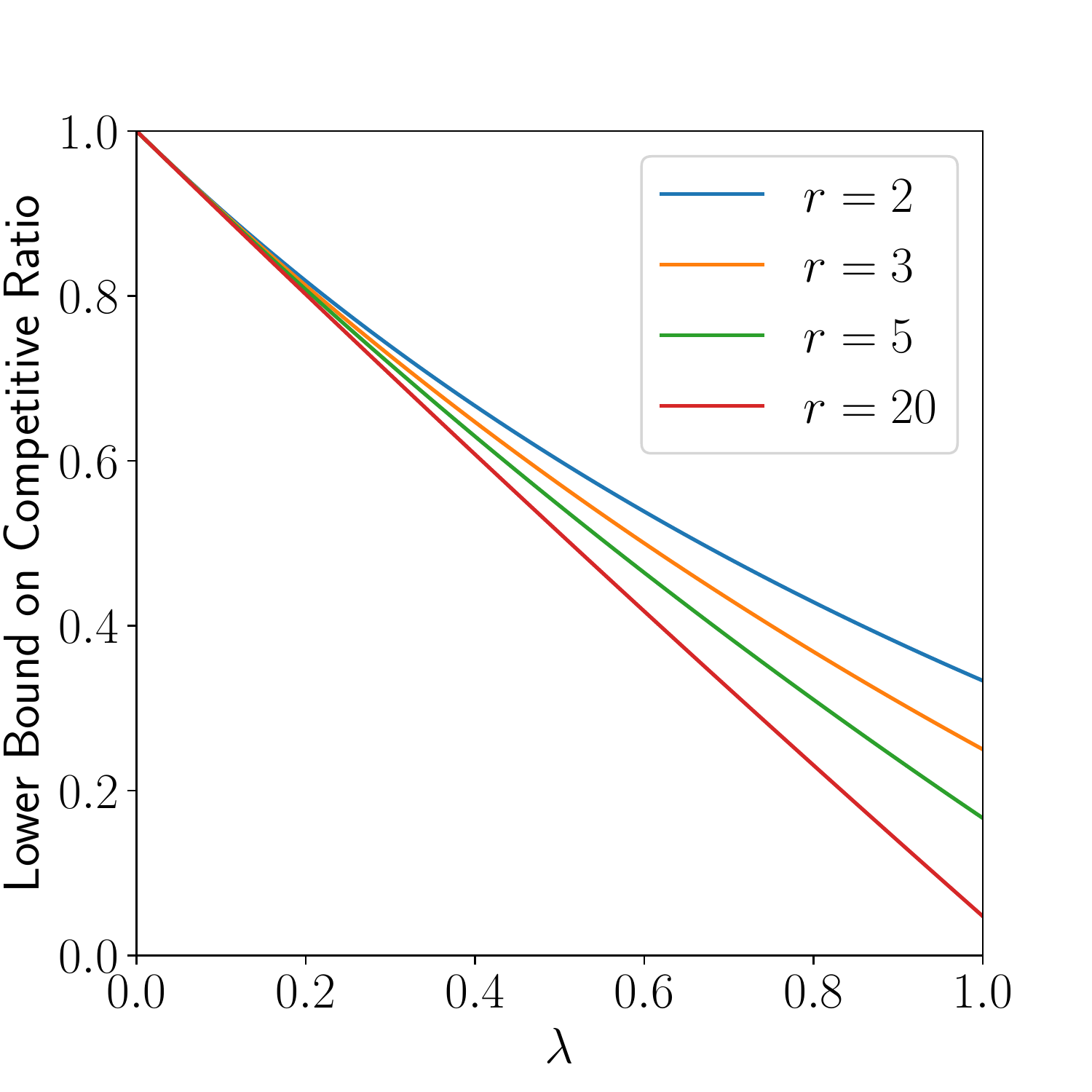}
    \caption{The lower bound \eqref{equ:strict-monotone-parallel-bounds-graph} on the optimal competitive ratio when the objective function has total curvature $\lambda$, as given in Theorem \ref{thm:best-parallel-strict-monotone-graph}.
    For $r = 2, 3, 5, 20$, we plot the lower bound against $\lambda$.
    Note that at $\lambda = 0$, the competitive ratio becomes 1, as expected from our intuition, and at $\lambda = 1$, the lower bound is $1/(r+1)$, which is consistent with \eqref{equ:parallel-exact-graph}.
    }
    \label{fig:plot-beta}
\end{figure}

\begin{definition}
Objective function $f : 2^S \to \mathbb{R}_{\ge 0}$ has \textit{total curvature} $\lambda \in (0,1)$ if $f(e \mid A) \ge (1-\lambda) f(e)$ for $e \in S$ and $A \subseteq S \setminus \{e\}$.
We denote the set of normalized, monotonic and submodular functions with total curvature $\lambda$ as $\mathcal{F}_\lambda$.
\end{definition}

We can extend competitive ratio defined in \eqref{def:comp-ratio-graph} and \eqref{def:comp-ratio-over-all-graph} to when they are subject to total curvature of $\lambda$:
\begin{equation}
\label{def:comp-ratio-graph-beta}
    \gamma_\lambda(G) = \inf_{f \in\mathcal{F}_\lambda, X} \gamma(f, X, G)
\end{equation}
\begin{equation}
\label{def:comp-ratio-over-all-graph-beta}
    \eta_\lambda(n, q) = \sup_{G \in \mathcal{G}_{n, q}} \gamma_\lambda(G)
\end{equation}

We can generalize the result of Lemma \ref{thm:david} to the total curvature setting:
\begin{lemma}
\label{thm:strict-monotone-lower}
Given any information graph $G$,
\begin{equation}
\label{equ:strict-monotone-lower}
    \frac{\alpha(G) - (\alpha(G)-1)\lambda}{\alpha(G)} \ge \gamma_\lambda(G) \ge \frac{\theta(G) - (\theta(G)-1)\lambda}{\theta(G) + \lambda}
\end{equation}
\end{lemma}

We then generalize Theorem \ref{thm:best-parallel-graph} to the total curvature setting and demonstrates that the iteration assignment \eqref{equ:complement-turan} is nearly optimal.
\begin{theorem}
\label{thm:best-parallel-strict-monotone-graph}
    Given a parallelized submodular maximization problem with $n$ agents and $q$ iterations, where the objective function has total curvature $\lambda$, then:
    \begin{equation}
    \label{equ:strict-monotone-parallel-bounds-graph}
        \frac{r - (r-1) \lambda}{r} \ge \eta_\lambda(n, q) \ge \frac{r - (r-1) \lambda}{r + \lambda}
    \end{equation}
    where $r = \lceil n / q \rceil$.
    Furthermore, the graph $G$ as defined in \eqref{equ:complement-turan} achieves the lower bound $\gamma_\lambda(G) \ge \frac{r - (r-1) \lambda}{r + \lambda}$.
\end{theorem}

If there is no parallelization (when $r=1$), \cite{conforti1984submodular} showed that the lower bound in \eqref{equ:strict-monotone-parallel-bounds-graph} is tight.

Figure \ref{fig:plot-beta} illustrates how the lower bound changes with respect to $\lambda$ at different $r$.
We can make some quick observations on how the lower bound changes with respect to $\lambda$.
In the limit of $\lambda \to 0$, the competitive ratio tends to 1, as expected from our intuition, and in the limit of $\lambda \to 1$, the lower bound tends to $1/(r+1)$, which is consistent with \eqref{equ:parallel-exact-graph}.
This confirms the intuition that for smaller $\lambda$, decisions have less ``overlap'' and as a result the greedy algorithm can perform closer to the optimal.
Also, as $r \to \infty$ both the lower and upper bounds converge to $1 - \lambda$.
So the lower and upper bounds are close to each other if $r$ is large (when there are a lot of parallelization).

For full proofs of the results in this section, please refer to Appendices \ref{sec:proof-strict-monotone-lower} and \ref{sec:complete-proof-parallel-beta}.

\section{Conclusion}
In this paper, we derived bounds on the competitive ratio of the parallelized greedy algorithm for both submodular objective functions and those with an additional property of total curvature.
We also provided the optimal design which achieves such bound and showed that a graph theoretic approach yields more effective parallelization that still achieves the same competitive ratio.

There are several directions of future research.
One possibility is to consider whether employing other structural properties on the objective functions can also improve the competitive ratio of greedy algorithm.
In particular, we are interested in properties that consider fixed number decisions at once because the total curvature property considers arbitrarily many decisions at once.
Another possible direction is to consider applying something other than the marginal contribution in making the greedy decisions.

\bibliographystyle{IEEEtran}
\bibliography{refs}

\appendix

\section{Proof of Theorem \ref{thm:best-parallel-graph}}
\label{sec:complete-proof-parallel}

We shall prove Theorem \ref{thm:best-parallel-graph} by applying Lemmas \ref{thm:parallel-indep-number}, \ref{thm:david} and \ref{thm:david-sibling}.

We first consider when $n \equiv 1 \pmod{q}$.
Combining Lemmas \ref{thm:parallel-indep-number} and \ref{thm:david} yields that $\gamma(G) \le 1/r$.
Now let $G$ be the graph described in \eqref{equ:weird-graph} and we can show that $\gamma(G) \ge 1/r$.
First note that for any $i < n$, $\overline{P}(G; i) = \lceil i / (r-1) \rceil$ and if $i = n$, $\overline{P}(G; i) = q$.
Hence this graph is in $\mathcal{G}_{n, q}$.
Let $T = \{i \mid i \le (r-1)(q-1) \}$, and note that for all $i \le n$, $\mathcal{N}_i \subseteq T$.
Then we have:
\begin{subequations}
\begin{align}
    f(\xopt) &\le f(\xopt, \xsol_T) \label{equ:weird-graphL1} \\
    &= f(\xsol_T) + \sum_{i=1}^n f(\xopt_i \mid \xopt_{1:i-1}, \xsol_T) \label{equ:weird-graphL2} \\
    &\le f(\xsol_T) + \sum_{i=1}^n f(\xopt_i \mid \xsol_{\mathcal{N}_i}) \label{equ:weird-graphL3} \\
    &= f(\xsol_T) + \sum_{j=1}^{r-1} \sum_{i=0}^{q-1} f(\xopt_{i(r-1)+j} \mid \xsol_{\mathcal{N}_i}) + f(\xopt_n \mid \xsol_T) \label{equ:weird-graphL4} \\
    &\le f(\xsol_T) + \sum_{j=1}^{r-1} \sum_{i=0}^{q-1} f(\xsol_{i(r-1)+j} \mid \xsol_{\mathcal{N}_i}) + f(\xsol_n \mid \xsol_T) \label{equ:weird-graphL5} \\
    &=  \sum_{j=1}^{r-1} f(\xsol_{\{j, j+(r-1), \dots, j+(r-1)(q-1)\}}) + f(\xsol_n, \xsol_T) \label{equ:weird-graphL6} \\
    &\le r f(\xsol) \label{equ:weird-graphL7}
\end{align}
\end{subequations}
where \eqref{equ:weird-graphL1} and \eqref{equ:weird-graphL7} follow from monotonicity, \eqref{equ:weird-graphL2} and \eqref{equ:weird-graphL6} follow from telescoping sums, \eqref{equ:weird-graphL3} follows from submodularity, \eqref{equ:weird-graphL4} rearranges the sum according to the graph structure as defined in \eqref{equ:weird-graph}, and \eqref{equ:weird-graphL5} follows from the definition of the greedy algorithm as stated in \eqref{def:general-greedy}.
From here we can conclude that $\eta(n, q) = 1/r$ when $n \equiv 1 \pmod{q}$.

Now we consider the case where $n \not\equiv 1 \pmod{q}$.
If $\alpha(G) > r$, then by Lemma \ref{thm:david}, $\gamma(G) \le 1/(r + 1)$.
If $\alpha(G) = r$, we suppose for the sake of contradiction that condition of Lemma \ref{thm:david-sibling} fails to hold.
Let $q' = \argmax_k |D_k| > 0$.
Then we need $|D_k| < r$ for any $k < q'$, otherwise we can pick $w$ be any agent in iteration $k+1$.
So we can have at most $(r-1)(q'-1) + r \le (r-1)q + 1$ agents.
But the condition $n \not\equiv 1 \pmod{q}$ implies that $n > q(r-1) + 1$, contradiction.
So by Lemma \ref{thm:david-sibling}, $\gamma(G) \le 1/(r + 1)$.

Now let $G$ be the graph described by \eqref{equ:complement-turan}.
For any $i \in n$, $\overline{P}(G; i) = \lceil i / r \rceil \le q$, hence this graph is in $\mathcal{G}_{n, q}$.
Also note that this graph consists of $r$ disjoint cliques, therefore $r = \alpha(G) = \theta(G)$.
By Lemma \ref{thm:david}, we have $\gamma(G) \ge 1/(r + 1)$, hence the equality is achieved. \hfill$\blacksquare$
\section{Proof of Lemma \ref{thm:strict-monotone-lower}}
\label{sec:proof-strict-monotone-lower}

We first show the lower bound on $\gamma_\lambda(G)$.
In this proof, we consider some minimum clique covering $\{K_1, K_2, \dots, K_\theta\}$ of $G$, where we can choose the $K$'s so they are disjoint.
Let $K(i)$ denotes the clique containing the $i$ node, and $\sigma_j = f(\xsol_{K_j})$.
We will upper bound $f(\xopt)$ in terms of the $\sigma$'s.
Then we express the quantity $\sigma_j/f(\xsol)$ as a concave function in $\sigma_j$ and using convexity to derive the final lower bound.

First, we need to resolve some technicalities so we have an appropriate objective function $f$ and action profile $X$.\footnote{The readers can safely ignore this paragraph without affecting their understanding of the rest of this proof.}
Given $(f \in \mathcal{F}_\lambda, X)$, we can assume that $X_{i} = \{\xsol_i, \xopt_i\}$ without affecting the competitive ratio.
We want $\xopt_i \neq \xsol_i$ for all $i \in N$, which ensures that all decisions are distinct.
Suppose not and for some $i \in N$, $\xopt_i = \xsol_i = u$ for some decision $u$.
Now we create a new decision $v$ and transform $(f, X)$ to $\left(\overline{f}, \overline{X}\right)$ so that $\overline{\xsol_j} = \xsol_j$, $\overline{\xopt_j} =\xopt_j$ for any $j \neq i$, $\overline{\xsol_i} = u$ and $\overline{\xopt_i} = v$.
And we define $\overline{f}$ as:
\begin{equation}
\label{equ:transformation}
\begin{cases}
    \overline{f}(u, v, A) = f(u, A) + (1 - \lambda) f(u) \\
    \overline{f}(u, A) = \overline{f}(v, A) = f(u, A) \\
    \overline{f}(A) = f(A)
\end{cases}
\end{equation}
for any $A \subseteq S \setminus X_i$.
Note that the above transformation does not affect the competitive ratio because our construction guarantees $\overline{f}\left(\overline{\xsol}\right) = f(\xsol)$ and $\overline{f}\left(\overline{\xopt}\right) = f(\xopt)$.
Also, from some simple algebra:
\begin{subequations}
\label{equ:after-transformation}
\begin{align}
    \overline{f}(u \mid v, A) = \overline{f}(v \mid u, A) &= (1-\lambda) f(u) \label{equ:after-transformationL1} \\
    \overline{f}(y \mid u, v, A) &= f(y \mid u, A) \label{equ:after-transformationL2}
\end{align}
\end{subequations}
where $A \subseteq S \setminus X_i$, $y \in S \setminus X_i$.
From \eqref{equ:after-transformation}, it is easy to verify that $(\overline{f}, \overline{X})$ satisfies all properties of submodular functions and has total curvature $\lambda$; but for brevity, we will not explicitly show them here.
Hence, for rest of the proof, we can safely assume that $\xopt_i \neq \xsol_i$ for all $i \in N$.

Now we bound $f(\xopt)$ through total curvature:
\begin{subequations}
\label{equ:cmb-opt-sol}
\begin{align}
    f(\xopt) + (1-\lambda) \sum_{j=1}^\theta \sigma_j
    &\le f(\xopt) + \sum_{j=1}^\theta \sum_{i \in K_j} (1-\lambda) f(\xsol_i) \label{equ:cmb-opt-solL1} \\
    &= f(\xopt) + \sum_{i=1}^n (1-\lambda) f(\xsol_i) \label{equ:cmb-opt-solL2} \\
    &\le f(\xopt) + \sum_{i=1}^n f(\xsol_i \mid \xsol_{1:i-1}, \xopt) \label{equ:cmb-opt-solL3} \\
    &= f(\xopt, \xsol) \label{equ:cmb-opt-solL4}
\end{align}
\end{subequations}
where \eqref{equ:cmb-opt-solL1} follows from submodularity, \eqref{equ:cmb-opt-solL2} follows from the fact that $K's$ are disjoint, \eqref{equ:cmb-opt-solL3} follows from total curvature, and \eqref{equ:cmb-opt-solL4} is a telescoping sum.

Then we bound $f(\xopt, \xsol)$ using properties of submodular functions and the greedy algorithm.
\begin{subequations}
\label{equ:telescope-clique}
\begin{align}
    f(\xopt, \xsol) &= f(\xsol) + \sum_{i=1}^n f(\xopt_i \mid \xopt_{1:i-1}, \xsol) \label{equ:telescope-cliqueL1} \\
    &\le f(\xsol) + \sum_{i=1}^n f(\xopt_i \mid \xsol_{\mathcal{N}_i}) \label{equ:telescope-cliqueL2} \\
    &\le f(\xsol) + \sum_{i=1}^n f(\xsol_i \mid \xsol_{\mathcal{N}_i}) \label{equ:telescope-cliqueL3} \\
    &\le f(\xsol) + \sum_{i=1}^n f(\xsol_i \mid \xsol_{\mathcal{N}_i} \cap K(i)) \label{equ:telescope-cliqueL4} \\
    &= f(\xsol) + \sum_{j=1}^\theta \sum_{i \in K_j} f(\xsol_i \mid \xsol_{\mathcal{N}_i} \cap K_j) \label{equ:telescope-cliqueL5} \\
    &= f(\xsol) + \sum_{j=1}^\theta \sigma_j \label{equ:telescope-cliqueL6}
\end{align}
\end{subequations}
where \eqref{equ:telescope-cliqueL1} and \eqref{equ:telescope-cliqueL6} are telescoping sums, \eqref{equ:telescope-cliqueL2} and \eqref{equ:telescope-cliqueL4} follow from submodularity, \eqref{equ:telescope-cliqueL5} follows from the fact that $K$'s are disjoint, and \eqref{equ:telescope-cliqueL3} follows from the greedy algorithm as defined in \eqref{def:general-greedy}.

Combining \eqref{equ:cmb-opt-sol} and \eqref{equ:telescope-clique}, we have that
\begin{equation}
f(\xopt) + (1-\lambda) \sum_{j=1}^\theta \sigma_j \le f(\xsol) + \sum_{j=1}^\theta \sigma_j
\end{equation}
Rearrange the terms yields:
\begin{equation}
\label{equ:bound-opt-clique}
    \frac{f(\xopt)}{f(\xsol)} \le 1 + \lambda \sum_{j=1}^\theta \frac{\sigma_j}{f(\xsol)}
\end{equation}
And we can upper bound the $\sigma$'s using total curvature in a manner similar to that of \eqref{equ:cmb-opt-sol}.
\begin{subequations}
\begin{align}
    \sigma_j + \sum_{\ell \neq j} (1 - \lambda) \sigma_\ell
    \le{}& \sigma_j + \sum_{i \not\in K_j} (1 - \lambda) f(\xsol_i) \label{equ:sum-sigmaL1} \\
    \le{}& \sigma_j + \sum_{i \not\in K_j} f(\xsol_i \mid \xsol_{1:i-1} \cup \xsol_{K_j}) \label{equ:sum-sigmaL2} \\
    ={}& f(\xsol) \label{equ:sum-sigmaL3}
\end{align}
\end{subequations}
where \eqref{equ:sum-sigmaL1} follows from submodularity, \eqref{equ:sum-sigmaL2} follows from total curvature, and \eqref{equ:sum-sigmaL3} follows from telescoping sum.

Without loss of generality, we impose the normalization that $\sum_{\ell=1}^\theta \sigma_\ell = \theta$.
Then for any $1 \le j \le \theta$,
\begin{equation}
\label{equ:upper-clique}
    \frac{\sigma_j}{f(\xsol)} \le \frac{\sigma_j}{(1-\lambda) \sum_{\ell=1}^\theta \sigma_\ell + \lambda \sigma_j} = \frac{\sigma_j}{(1-\lambda) \theta + \lambda \sigma_j}
\end{equation}

We can denote the right hand side of \eqref{equ:upper-clique} by the function $g(x) = x / ((1-\lambda) x + \lambda x)$.
Note that $g$ is a concave function.
We then apply the Jensen's inequality:
\begin{lemma}[Jensen]
For a concave function $g$ and values $y_1, \dots, y_m$ in its domain, then
\begin{equation}
    \frac{1}{m} \sum_{i=1}^m g(y_i) \le g\left(\frac{\sum_{i=1}^m y_i}{m}\right)
\end{equation}
And the equality holds if and only if $y_1 = \cdots = y_m$ or when $g$ is linear.
\end{lemma}

We substitute our normalization into \eqref{equ:upper-clique} and combine it with \eqref{equ:bound-opt-clique}:
\begin{equation}
    \frac{1}{\gamma_\lambda(G)} = \frac{f(\xopt)}{f(\xsol)} \le 1 + \frac{\lambda \theta}{(1-\lambda) \theta + \lambda} = \frac{\theta + \lambda}{\theta - (\theta-1)\lambda}
\end{equation}
Hence we derived the lower bound.

Now we show the upper bound on $\gamma_\lambda(G)$.
Let $I$ be a maximum independent set of $G$ and $\alpha = |I|$.
Define two sets of distinct decisions $U = \{u_i, u_2, \dots, u_\alpha\}$ and $V = \{v_1, v_2, \dots, v_\alpha\}.$
Consider $S = U \cup V$ and action profile determined by:
\begin{equation}
X_{i} = 
\begin{cases}
    \{u_i. v_i\} & \text{if } i \in I \\
    \emptyset & \text{otherwise}
\end{cases}
\end{equation}
Define the objective function $f$ so that for any $x \in X$ (with $x$ interpreted as subset of $S$),
\begin{equation}
f(x) = \min(1, |x \cap U|)\lambda + |x \cap U| (1-\lambda) + |x \cap V|
\end{equation}

Note that through this construction, $f$ is a submodular function and has total curvature $\lambda$.
Also, $f$ has the following properties for any $i \in I$:
\begin{enumerate}
    \item $f(u_i) = f(v_i) = 1$.
    \item $f(u_i \mid x_{\mathcal{N}_i}) = f(u_i)$ for any $x \in X$ because by definition, we have $\mathcal{N}_i \cap I = \emptyset$.
    \item $f(v_i | B) = f(v_i)$ for any $B \subseteq S \setminus \{v_i\}$
\end{enumerate}

From these properties, the agents in $I$ are equally incentivized to pick either option.
In one of the greedy solutions, the action would be the $u$'s.
And in the optimal solution, the action would be the $v$'s.
This results in $f(\xsol) \le \lambda + \alpha(1 - \lambda)$ and $f(\xopt) = \alpha$, hence $\gamma_\lambda \le \frac{\alpha - (\alpha-1)\lambda}{\alpha}$.
\hfill$\blacksquare$
\section{Proof of Theorem \ref{thm:best-parallel-strict-monotone-graph}}
\label{sec:complete-proof-parallel-beta}

First we note that the function $g(x) = (x - (x-1) \lambda) / x$ is decreasing, and the upper bound in \eqref{equ:strict-monotone-lower} is equal to $g(\alpha(G))$.
The upper bound in \eqref{equ:strict-monotone-parallel-bounds-graph} then follows from combining Lemma \ref{thm:parallel-indep-number} and the upper bound in Lemma \ref{thm:strict-monotone-lower}.

Consider the graph $G$ described by \eqref{equ:complement-turan}; as we already argued previously in Appendix \ref{sec:complete-proof-parallel}, $r = \alpha(G) = \theta(G)$ and $G \in \mathcal{G}_{n, q}$.
Hence, from the lower bound in Lemma \ref{thm:strict-monotone-lower}, we conclude that this graph achieves the lower bound in \eqref{equ:strict-monotone-parallel-bounds-graph}, so we are done. \hfill$\blacksquare$


\end{document}